\documentclass[10pt,journal]{IEEEtran}
\usepackage{cite}
\usepackage{amsmath,amssymb,amsfonts}
\usepackage{algorithmic}
\usepackage{graphicx}
\usepackage{textcomp}
\usepackage{xcolor}
\usepackage{setspace}
\usepackage{latexsym}
\usepackage{graphicx}
\usepackage{amssymb}
\usepackage{epsfig}
\usepackage{float}
\usepackage{caption}
\usepackage{subfigure}
\usepackage{bm}
\usepackage{hyperref}

\usepackage{geometry}
\def\BibTeX{{\rm B\kern-.05em{\sc i\kern-.025em b}\kern-.08em
    T\kern-.1667em\lower.7ex\hbox{E}\kern-.125emX}}
\begin{document}

\title{Reconfigurable Intelligent Surface-induced Randomness for mmWave Key Generation} % in Static Environments

 	\author{Shubo~Yang, Han~Han, Yihong~Liu,~\IEEEmembership{Graduate~Student~Member,~IEEE}, Weisi~Guo,~\IEEEmembership{Senior~Member,~IEEE}, Zhibo Pang, and Lei~Zhang,~\IEEEmembership{Senior~Member,~IEEE} 
	\thanks{ S.~Yang is with the Glasgow College, University of Glasgow (e-mail: 2429400y@student.gla.ac.uk). H.~Han is with Electrical and Computer Engineering Department, University of Toronto, ON M5S 3G4, Canada (e-mail: johnny.han@mail.utoronto.ca). Y.~Liu and L.~Zhang are with James Watt School of Engineering, University of Glasgow, Glasgow G12 8QQ, U.K. (e-mail: y.liu.6@research.gla.ac.uk; lei.zhang@glasgow.ac.uk). W.~Guo is with School of Aerospace, Transport and Manufacturing, Cranfield University, MK43 0AL Cranfield, U.K. (e-mail: weisi.guo@cranfield.ac.uk). Z. Pang is with Department of Automation Technology, ABB Corporate Research Sweden, Vasteras, Sweden, and Department of Intelligent Systems, Royal Institute of Technology (KTH), Stockholm, Sweden. (e-mail: pang.zhibo@se.abb.com, zhibo@kth.se).
		
	Funding information:
	Zhibo Pang’s work is partly funded by the Swedish Foundation for Strategic Research (SSF) through the project APR20-0023.
	} }
%\vspace{-5mm}}

\maketitle
\begin{abstract}
	Secret key generation in physical layer security exploits the unpredictable random nature of wireless channels. The millimeter-wave (mmWave) channels have limited multipath and channel randomness in static environments. In this paper, for mmWave secret key generation of physical layer security, we use a reconfigurable intelligent surface (RIS) to induce randomness directly in wireless environments, without adding complexity to transceivers. We consider RIS to have continuous individual phase shifts (CIPS) and derive the RIS-assisted reflection channel distribution with its parameters. Then, we propose continuous group phase shifts (CGPS) to increase the randomness specifically at legal parties. Since the continuous phase shifts are expensive to implement, we analyze discrete individual phase shifts (DIPS) and derive the corresponding channel distribution, which is dependent on the quantization bit. We then derive the secret key rate (SKR) to evaluate the randomness performance. With the simulation results verifying the analytical results, this work explains the mathematical principles and lays a foundation for future mmWave evaluation and optimization of artificial channel randomness.
%	\vspace{-2mm}
\end{abstract}

\begin{IEEEkeywords}
	Physical layer security, secret key generation, reconfigurable intelligent surface, intelligent reflecting surface.
%	\vspace{-2mm}
\end{IEEEkeywords}
\section{Introduction}\label{Introduction}
%\vspace{-2mm}
\IEEEPARstart{W}{ireless} networks are becoming ubiquitous nowadays and in the future Internet of Things (IoT) systems. However, their broadcast nature makes them vulnerable to malicious attacks. Classic encryption schemes, such as advanced encryption standard and public key cryptography, are dependent on cryptography computation techniques \cite{8715341}. The applications of classic schemes to IoT devices and wireless sensor networks (WSNs) bring challenges, since the devices and sensor nodes have small sizes and limited computational capability. Thus, extensive research is carried out in secret key generation of physical layer security, where the legitimate users extract keys from their correlated observations of the reciprocal channel in a lightweight manner \cite{7557048}. The correlation of channels makes it possible to generate keys without key exchange, and the dynamic uniqueness of the channel prevents eavesdroppers from mimicking. While in dynamic environments the movements of users or objects are sufficient to produce randomness, the randomness is limited in static environments, such as in open terrain with no moving objects. Besides, the millimeter wave (mmWave) communication is envisioned as a significant technology for the fifth generation (5G) networks and beyonds \cite{10.1007/978-3-319-99007-1_43}. Thus, the security in mmWave static environments needs to be researched. 

The newly developed reconfigurable intelligent surface (RIS), also known as Intelligent Reflecting Surface (IRS), has the potential to produce artificial randomness in mmWave static environments. RIS is a two-dimensional surface consisting of a large number of passive low-cost reflecting elements \cite{9326394}. Each scattering element of RIS is independently capable of altering the amplitude and/or phase of the incident signals. Additionally, RIS becomes more important in high-frequency band communications, e.g., mmWave and THz communications that have severe coverage issues. The existing RIS applications mainly target indoor static scenarios \cite{9122596}. Therefore, RIS can be easily incorporated in mmWave static environments. By adding RIS, artificial randomness can be produced. The randomness does not rely on dynamic environments and is produced directly in the channel, without needing increased transceiver costs.

Research has been done on RIS-assisted key generation. Random shifting RIS is applied to increase the secure transmission rate, and the time allocation for key generation and transmission is designed \cite{2020arXiv201014268J}. RIS with discrete phase shifts is adopted to generate secret keys, and the secret key rate (SKR) is derived \cite{9442833}. The practical implementation of using RIS in the OFDM system is conducted \cite{9569556}. However, most existing literature focuses on sub-6 GHz systems and models on Gaussian channels. The mmWave channels have a poor scattering nature and exhibit limited multipath, so they may not conform to Gaussian channels. Moreover, the channel distribution resulted from RIS phase shifts and element number is still unknown. Besides, most previous works assume each RIS element phase shift is independent and identically distributed (i.i.d), without changing phase shift distributions to improve randomness. 

Therefore, in this paper, the RIS-assisted mmWave key generation is proposed. We model RIS-induced randomness in mmWave key generation, and we focus on the fundamental analytical derivations of channel distribution with random RIS weights. For the random weights, we consider applying continuous individual phase shifts (CIPS) on each element and continuous group phase shifts (CGPS) on elements in groups to produce higher randomness. In addition, we consider channels for both continuous\footnote{The continuous RIS weights are difficult to implement. This is because more levels of weight phase shift result in more costs, which is not scalable to a large number of elements \cite{8930608}. The continuous weights' performance is the upper limit for discrete weights when weights' quantization bits approach infinity.} and discrete phase shifts and compare their performance. To summarize, the main contributions of this paper are as follows.
%  Thus, 
\begin{itemize}
	\item The RIS-induced channel distribution and its parameters are derived, given RIS is a uniform rectangular array (URA) and the elements have CIPS. As the result, an artificial Rayleigh/Rician fading is induced directly in the environments.
	\item To increase the amount of induced randomness, CGPS is proposed and the channel is derived. The channel variance for legal parties increases, and there is a tradeoff between group number and group size. The discrete individual phase shifts (DIPS) is also discussed based on quantization bits.
	\item The SKR is derived for CIPS, CGPS, and DIPS, to evaluate the performance of artificial randomness.
\end{itemize}

\emph{Notations:} Bold-faced letters are used to denote matrix or column vectors, while lightfaced letters are used to denote scalar quantities. Superscripts $ (\cdot)^T $, $ (\cdot)^* $, and $ (\cdot)^H $ represent the transpose, conjugate and conjugate transpose operations, respectively. $ \odot $ denotes the point-wise multiplication. We use the notations shown in Table \ref{table1} in this paper. 
%\vspace{-2mm}
\begin{table}[h] 
	\centering 
	\caption{Notations}
	\label{table1}
	\resizebox{1\columnwidth}{!}{
		\begin{tabular}{|c|l|}  
			\hline
			Symbol & Definition\\ \hline  
			$ \lambda $ & Wavelength\\
			$ k $ & Wave number, where $ k = \frac{2\pi}{\lambda}$\\
			$ d $ & Element spacing in RIS\\
			$ M $ & Element number of the RIS\\
			$ m $ & An integer in range $[1,M]$ \\
			$\phi_m$ & The phase shift of the $m$-th RIS element weight \\
			$\psi$  & The incident or reflected azimuth angle\\
			$\theta$  & The incident or reflected elevation angle\\
			$ B $ & RIS discrete weight quantization bit\\
			$\bm{R}$ & The covariance between the real and imaginary parts of a complex Gaussian distribution\\
			$R_s$ & Secret key rate \\
			$U(a,b)$ & Indicate a uniform distribution on interval $(a,b)$ \\
			$N(\mu,\sigma^2)$ & Indicate a Gaussian distribution with mean $\mu$ and variance $\sigma^2$\\
			$h_{(\cdot)}$ & Indicate the channel\\
			$(\cdot)_x, (\cdot)_y$ & Indicate physical quantities on $x$-axis and $y$-axis, respectively\\
			$(\cdot)_{real}, (\cdot)_{imag}$ & Indicate quantities of the real and imaginary parts, respectively\\ % in a complex number
			$(\cdot)_{cos}, (\cdot)_{sin}$ & Indicate quantities of cosine and sine functions, respectively\\
			$(\cdot)_{i}, (\cdot)_{o}$ & Indicate incident and reflected angles, respectively\\
			$(\cdot)_{CI}$ & Indicate quantities when RIS weights have continuous individual phase shifts\\
			$(\cdot)_{CG}$ & Indicate quantities when RIS weights have continuous group phase shifts\\
			$(\cdot)_{DI}$ & Indicate quantities when RIS weights have discrete individual phase shifts\\
%			$\Re(\cdot)$ & Indicate the real part\\
%			$\Im(\cdot)$ & Indicate the imaginary part\\
			\hline
		\end{tabular}
	}
\end{table}

\section{System Model}\label{System Model}
Consider an RIS-assisted wireless communication between legitimate parties, Alice and Bob. There is also an eavesdropper Eve, who passively listens to the communication. Assume all the parties are operated under a mmWave static environment. We consider the narrowband line-of-sight (LOS) transmission in the far field between Alice/Bob and RIS. 
\begin{figure}[htb]
	\centering
	\includegraphics[width=80mm,height=40mm]{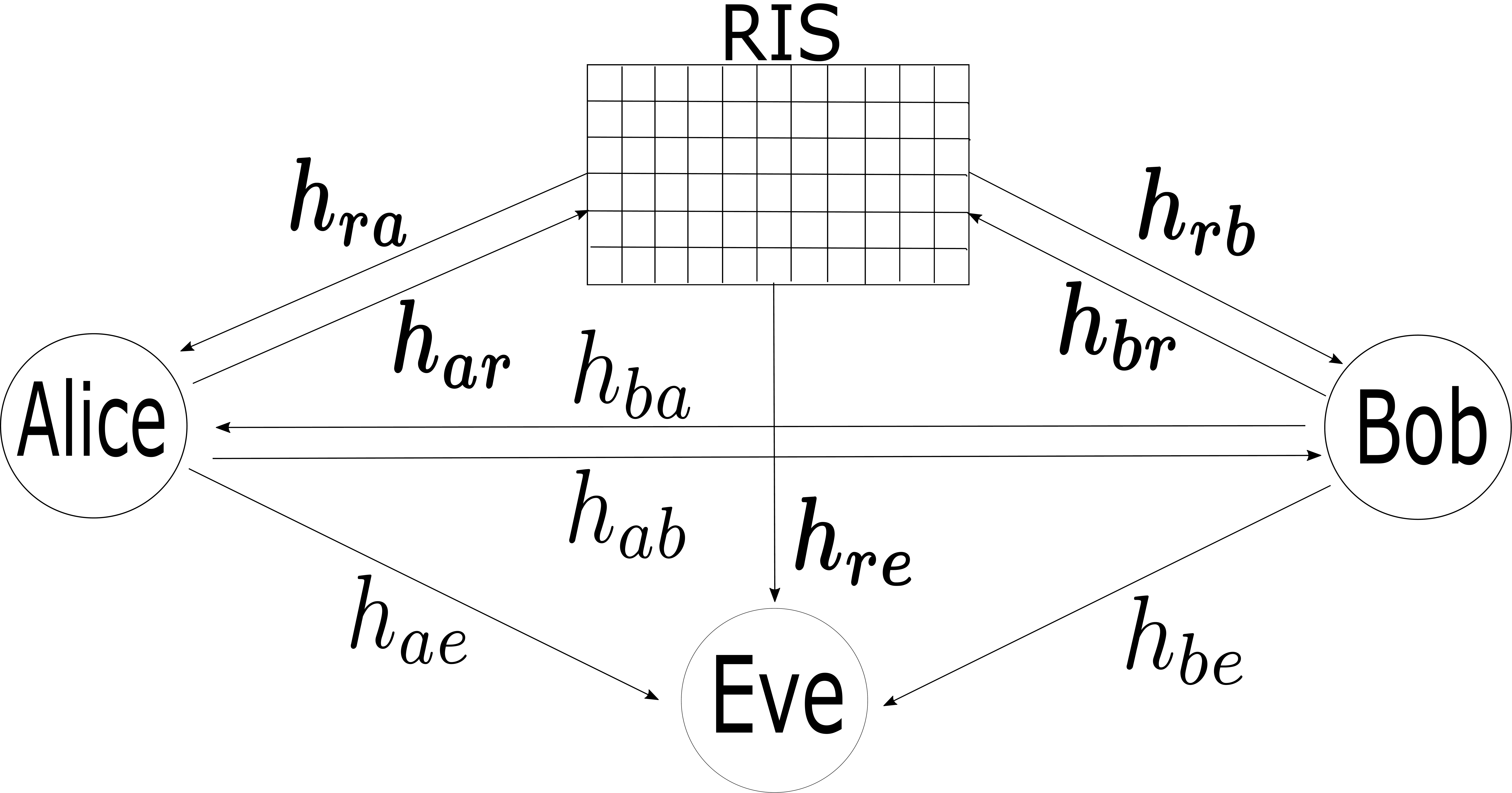}
	\caption{RIS-assisted secret key generation model.}
	\label{model}
	\vspace{-4mm}
\end{figure}
This assumption can be justified since the multipath effect is constrained and the high path loss makes non-line-of-sight (NLOS) paths' power small. The RIS provides cascaded LOS paths. Additionally, the path loss is a constant related to distance and can be easily predicted by eavesdropper \cite{9442833}. Since the transmission channel between Alice and Bob is reciprocal, we only consider the situation where Alice is transmitting signals to Bob to avoid redundancy. 

As shown in Fig. \ref{model}. Alice and Bob aim at establishing a random secret cryptographic key based on their reciprocal communication channel and reducing the key leakage to Eve. They measure the common channel through their exchanged signals and generate keys using parameters of the received signals, e.g. channel state information (CSI) and received signal strength (RSS). When Alice transmits signal $s$, the received signal at Bob can be expressed as
\begin{equation}
\label{received_signal}
y=(h_{ab}+\bm{h_{rb}^H} \bm{W} \bm{h_{ar}})s + z \ ,
\end{equation}
where $h_{ab} \in \mathbb{C}^{1 \times 1}$, $\bm{h_{ar}} \in \mathbb{C}^{M \times 1}$, $\bm{h_{rb}} \in \mathbb{C}^{M \times 1}$ represent the direct channel between Alice and Bob, the channel between Alice and RIS, and the channel between RIS and Bob, $z \sim \mathcal{CN}(0,\sigma_z^2)$ is the additive white Gaussian noise (AWGN), and $\bm{W}$ is the diagonal weight matrix with each entity on the diagonal being the RIS weight of each surface element. The weight of the $m$-th surface element is represented as $e^{j\phi_m}$.

To generate shared keys in static environments and reduce the key leakage to Eve, it is crucial to increase the channel randomness through RIS. As in (\ref{received_signal}), though $\bm{h_{ar}}$ and $\bm{h_{rb}}$ are static in mmWave static environments, when RIS weights $\bm{W}$ vary, $\bm{h_{rb}^H} \bm{W} \bm{h_{ar}}$ becomes dynamic. Thus, RIS becomes the major source of channel randomness through varying the reflection channel $\widetilde{H}=\bm{h_{rb}^H} \bm{W} \bm{h_{ar}}$. In this case, we focus on $\widetilde{H}$, which can be expressed using steering vectors and weight vector \cite{a1} as
%\vspace{-2mm}
\begin{equation}
\label{H_steer}
\widetilde{H}=\bm{w}^H \cdot [\bm{a}(\bm{\Omega}_{i}) \odot \bm{a}(\bm{\Omega}_{o})] \ ,
%\vspace{-2mm}
\end{equation}
where $\bm{w}$ is the weight vector with each entity being the diagonal entity of $\bm{W}$, i.e., weight of each surface element. Besides, the steering vector of the channels between Alice and RIS and between RIS and Bob are
\begin{equation}
\bm{a}(\bm{\Omega}_{i}) = [a(\bm{\Omega}_{i,1}),\cdots,a(\bm{\Omega}_{i,m}), \cdots,a(\bm{\Omega}_{i,M})]^T 
\end{equation}
\begin{equation}
\bm{a}(\bm{\Omega}_{o}) = [a(\bm{\Omega}_{o,1}),\cdots,a(\bm{\Omega}_{o,m}), \cdots, a(\bm{\Omega}_{o,M})]^T \ ,
\end{equation}
%	\in \mathbb{C}^{M\times 1}
where $ \bm{\Omega}_{i,m} $ and $\bm{\Omega}_{o,m} $ are the terms characterizing the incident spatial information from Alice at RIS $m$-th element and the reflected spatial information from $m$-th element to Bob, respectively. Since the environment is static, $\bm{\Omega}_{i}$ and $\bm{\Omega}_{o}$ are fixed, and $\widetilde{H}$ becomes a function of solely $\bm{w}$. Note that averaging $\widetilde{H}$ derived from all possible pairs of $\bm{\Omega}_{i}$ and $\bm{\Omega}_{o}$ provides an estimation of the overall channel distribution. 

\section{RIS-Assisted Key Generation} 
In this section, we first derive the probability density function (p.d.f) of reflection channel $\widetilde{H}$ with a given random RIS having CIPS and CGPS. Individual shifting means that the RIS phase shifts $\lbrace \bm{\phi}=[\phi_1, \cdots,\phi_m,\cdots,\phi_M], m \in [1,M] \rbrace$ are i.i.d. Group shifting means the elements phase shifts in a group are first implemented maximum ratio transmission (MRT) and then added a random phase shift. Next, we consider DIPS. The channel for 1-bit is found different from other quantization bits. Since group phase shifts need MRT to compensate for the phase difference caused by element distance, the discrete group phase shifts (DGPS) need multiple levels to realize the compensation. Implementing multiple-level discrete phase shifts is expensive compared to low-level phase shifts. Thus, we only discuss CGPS, and DGPS will be our future work.

Note that a uniform linear array (ULA) can be considered a special case of URA while URA is a typical planar array structure. Thus, we consider RIS in URA configuration, which can be further extended into other complex RIS shapes. 

%\vspace{-2mm}
\subsection{Continuous Individual Phase Shifts} \label{continuous}
Consider a URA RIS with $M_x$ and $M_y$ elements equally spaced on x-axis and y-axis, respectively, where the total element number is $M=M_x \cdot M_y$. The incident and reflected angles are described by $\bm{\Omega}_{i}=(\psi_{i},\theta_{i})$ and $\bm{\Omega}_{o}=(\psi_{o},\theta_{o})$. Assume the i.i.d RIS weight phase shifts $\bm{\phi}$ are continuous uniformly distribution $\phi_m \sim U(0,2\pi)$. Then, the multiplication of steering vectors in (\ref{H_steer}) with elements spacing along x-axis $d_x$ and y-axis $d_y$ can be expressed as
\begin{equation} 
%	\vspace{-2mm}
\begin{aligned}
\bm{a}(\bm{\Omega}_{i})  \odot  \bm{a}(\bm{\Omega}_{o})  =  & [1,  \cdots, 
e^{j(\xi_x \cdot (m_x-1) +\xi_y \cdot (m_y-1))}, \\
&  \cdots ,  e^{j(\xi_x \cdot (M_x-1)  +\xi_y \cdot (M_y-1))}]^T  \   ,
\end{aligned}
\end{equation}
where $ m_x \in [1,M_x] $ and $ m_y \in [1,M_y] $ are integers \cite{9301343}. Additionally, $k=\frac{2\pi}{\lambda}$ is the wave number and
\begin{equation}
\xi_x = kd_x (\cos\psi_{i} \sin\theta_{i} + \cos\psi_{o} \sin\theta_{o}) \ ,
\end{equation}
%	\vspace{-1mm}
\begin{equation}
\xi_y = kd_y(\sin\psi_{i} \sin\theta_{i}+\sin\psi_{o} \sin\theta_{o}) \ .
\end{equation}
Consequently, the reflection channel following (\ref{H_steer}) can be constructed as 
\begin{equation}
\label{H}
\begin{aligned}
\widetilde{H} \! \! = &1+\cdots+	\! e^{j[\phi_{m}+(\xi_x \cdot (m_x-1) +\xi_y \cdot (m_y-1))]} \\
&+\cdots+ e^{j[\phi_M+(\xi_x \cdot (M_x-1) +\xi_y \cdot (M_y-1))]} \\
=& \sum_{m_y=1}^{M_y} \sum_{m_x=1}^{M_x}  e^{j(\phi_m+\alpha)} \\
=& \sum_{m_y=1}^{M_y} \sum_{m_x=1}^{M_x} \cos(\phi_m+\alpha) + 
j \sum_{m_y=1}^{M_y} \sum_{m_x=1}^{M_x} \sin(\phi_m+\alpha) \ ,
\end{aligned}
\vspace{-1mm}
\end{equation}
\vspace{-1mm}
where
\begin{equation}
m=(m_y-1) \cdot M_x +m_x \ , \ \alpha=(m_x-1)\cdot \xi_x+ (m_y-1) \cdot \xi_y \ .
\end{equation}
According to the central limit theorem (CLT), $\widetilde{H}$ converges to a complex Gaussian distribution when $M$ is large enough. This assumption could be justified since a practical RIS usually has an extremely large number of elements, e.g. more than tens of elements \cite{9122596}. Therefore, the real and imaginary parts both converge to Gaussian distributions that are fully determined by means and variances. The mean and variance of the real part of $\widetilde{H}_{CI}$ can be expressed as
\begin{equation}
\label{mu_real}
\mu_{real,CI} \! = \! \sum_{m_y=1}^{M_y} \! \! \sum_{m_x=1}^{M_x} \! \! \mu_{cos,CI}  \ , \  \sigma^2_{real,CI} \! = \! \! \sum_{m_y=1}^{M_y} \! \! \sum_{m_x=1}^{M_x} \! \! \sigma^2_{cos,CI} \ ,
\end{equation}
where 
\begin{equation}
\label{mu_cos}
\mu_{cos,CI}=\int_0^{2\pi}\cos(\phi_m  +   \alpha) f(\phi_m) d\phi_m = 0 \ .
\end{equation}
\begin{equation}
\label{var_cos}
\sigma^2_{cos,CI} \! \! = \! \! \int_0^{2\pi} \! \! \! \! \cos^2(\phi_m  +   \alpha) f(\phi_m) d\phi_m -\mu_{cos,CI }^2 \! = \! \frac{1}{2}.
%\vspace{-2mm}
\end{equation}
%Since $\mu_{cos,CI}$ and $\sigma^2_{cos,CI}$ remain constant, they are independent of the values of $\lbrace m_x, m_y \rbrace$ and $\lbrace \xi_x, \xi_y \rbrace$. 
The imaginary part can be derived similar to the real part. By substituting (\ref{mu_cos}) (\ref{var_cos}) into (\ref{mu_real}), the channel real and imaginary parts are
\begin{equation}
	\Re{(\tilde{H}_{CI})} \sim N(0,\frac{M}{2}) \ ,  \ 
	\Im{(\tilde{H}_{CI})} \sim N(0,\frac{M}{2}) \ .
	\label{norm}
\end{equation}
Since the real and imaginary parts of $\widetilde{H}_{CI}$ constitute a joint Gaussian distribution and the covariance of them equals 0, the real and imaginary parts are independent. The distribution of $\widetilde{H}_{CI}$ is also independent of $\bm{\Omega}_{i}$ and $\bm{\Omega}_{o}$.
With any pair of incident and reflected angles, the mean of $\widetilde{H}_{CI}$ is $\mu_{CI}=0$ while the variance of $\widetilde{H}_{CI}$ can be calculated as 
\begin{equation}
%\vspace*{-1mm}
\label{var_patt}
\sigma^2_{CI}=\mathbb{E}[(\widetilde{H}_{CI}-\mu_{CI})(\widetilde{H}_{CI}-\mu_{CI})^*]=M \ .
\end{equation}
Due to the channel parameters CSI and RSS are normally used to generate keys, the magnitude and phase distributions are also important. According to the joint Gaussian distribution, the result magnitude p.d.f and phase p.d.f of $\widetilde{H}_{CI}$ are Rayleigh and uniform distributions. When the direct path and reflection path are both considered, the direct path adds a non-zero mean, and the induced Rayleigh fading will turn into Rician fading. Hence, the CIPS RIS produces randomness by inducing an artificial Rayleigh/Rician fading in the environments. 

\subsection{Continuous Group Phase Shifts}
Different from the i.i.d elements in CIPS, group shifting means first using MRT to achieve maximum transmission rate at Alice and Bob, then shifting the phase randomly. This induces more randomness at Alice and Bob.

The total elements $M$ is divided into groups of $q$ elements. There are $N = \lfloor \frac{M}{q} \rfloor $ groups in total. The $n$-th group has the added random phase $\phi_{n,rand} \sim U(0,2\pi)$. When the $m$-th element is in $n$-th group, the phase shift for $m$-th element is
\begin{equation}
	\phi_{m,CG}=\phi_{m,MRT} + \phi_{n,rand} \ ,
\end{equation}
where 
\begin{equation}
	\phi_{m,MRT} = \frac{[\bm{a}(\bm{\Omega}_{i,m}) \cdot \bm{a}(\bm{\Omega}_{o,m})]^H}{||\bm{a}(\bm{\Omega}_{i,m}) \cdot \bm{a}(\bm{\Omega}_{o,m})||} \ .
\end{equation}
The $\phi_{m,MRT}$ compensates the steering vector phase difference of elements in the same group $n$ to achieve larger signal strength. The random phase shift $\phi_{n,rand}$ produces the randomness. Then, according to the channel in (\ref{H}), the channel for CGPS can be expressed as 
\begin{equation}
\widetilde{H}_{CG} = \sum_{n=1}^{N} q \cos(\phi_{n,rand}) + \ j \sum_{n=1}^{N} q \sin(\phi_{n,rand}) \ .
\end{equation}
When RIS has a large number of elements, the number of groups $N$ can be large when given a proper value of $q$. Then, $\widetilde{H}_{CG}$ converges to a complex Gaussian distribution. Similarly to Section \ref{continuous}, the mean $\mu_{real,CG}=0$ and the variance of the real part can be expressed as
\begin{equation}
\begin{aligned}
\sigma^2_{real,CG}=\sum_{n=1}^{N} \sigma^2[q \cos(\phi_{n,rand})] = \frac{Nq^2}{2} \ .
\end{aligned}
\end{equation}
% = \sum_{n=1}^{N} \mathbb{E}[q\cos(\phi_{n,random})] = 0 \ , \\
Thus, the channel distribution for CGPS RIS is
\begin{equation}
\Re{(\tilde{H}_{CG})} \sim N(0,\frac{Nq^2}{2}) \ , 
\Im{(\tilde{H}_{CG})} \sim N(0,\frac{Nq^2}{2}) \ .
\end{equation}
The channel variance can be then calculated as
\begin{equation}
	\sigma^2_{CG}=Nq^2 \ . 
	\label{var_patt_group}
\end{equation}
There is a tradeoff between $N=\lfloor \frac{M}{q} \rfloor$ and $q$. When $q$ is large the variance increases quadratically, but $N$ may become too small to use CLT. The Gaussian distribution is the distribution that maximizes the entropy at a given variance. The CGPS variance $\sigma^2_{CG}=Nq^2$ can achieve larger randomness than the CIPS at Alice and Bob.

\vspace{-2mm}
\subsection{Discrete Individual Phase Shifts}
\label{discrete} 
Since continuous phase shifts are expensive to implement, discrete phase shifts RIS need be researched. Assume RIS has discrete weight phase shifts $\bm{\phi}$ with quantization bit $B$. $\bm{\phi}$ is i.i.d uniformly on discrete values $\lbrace 0,\frac{2\pi}{2^B},\cdots,\frac{2\pi(2^B-1)}{2^B} \rbrace$. When $B$ approaches to infinity, the channel distribution approaches to the distribution assisted by RIS with continuous phase shifts. Similar to the CIPS, the DIPS reflection channel can be expanded as in (\ref{H}) and (\ref{mu_real}). The mean and variance of the real part $\mu_{cos,DI}$ and $\sigma^2_{cos,DI}$ are expressed as
\begin{equation}
	\mu_{cos,DI} =\mathbb{E} \lbrace \cos(\phi_m  +   \alpha) \rbrace = 0 \ , 
\end{equation}
\begin{equation}
\begin{aligned}
\label{mu_cos_D}
\!\!\sigma^2_{cos,DI}  \! \!
&=  \! \mathbb{E} \lbrace  \cos^2(\phi_m  +   \alpha)\rbrace -\mu_{cos,DI}^2 \\
& = \! \frac{1}{2}+\frac{1}{2} \mathbb{E}\lbrace \cos(2\phi_{m}+2\alpha)\rbrace \\
& = \! \! \frac{1}{2}\! \! + \! \! \frac{1}{2}\cos(2\alpha)\mathbb{E}\lbrace \! \cos(2\phi_{m}) \! \rbrace \!- \! \frac{1}{2}\sin(2\alpha)\mathbb{E}\lbrace \! \sin(2\phi_{m}) \! \rbrace \! \ \! ,
\end{aligned}
\end{equation}
where the means of $\cos(2\phi_{m})$ and $\sin(2\phi_{m})$ determine $\sigma^2_{cos,DI}$. According to quantization bits $B$, there are two cases of $\sigma^2_{cos,DI}$.

\subsubsection{$B \geq 2$} 
$\mathbb{E}\lbrace\cos(2\phi_{m})\rbrace=0$ and $\mathbb{E}\lbrace\sin(2\phi_{m})\rbrace=0$, which leads to $\sigma^2_{cos,DI}=\frac{1}{2}$. Therefore, same as CIPS, the channel is a complex Gaussian random variable with real and imaginary parts being same as in (\ref{norm}), that are, $\Re{(\tilde{H}_{DI})} \sim N(0,\frac{M}{2}) $ and $\Im{(\tilde{H}_{DI})} \sim N(0,\frac{M}{2}) $. The variance is the same as in (\ref{var_patt}) $\sigma^2_{DI} = M$.
%$\mu_{real,D}=\mu_{imag,D}=0$ and $\sigma^2_{real,D}=\sigma^2_{imag,D}=\frac{M}{2}$

\subsubsection{$B =1$} 
$\mu_{real,DI}=\mu_{imag,DI}=0$. Different from $B \geq 2$ case, since $\phi_{m}$ only takes values $\lbrace 0, \pi \rbrace$, $\mathbb{E}\lbrace\cos(2\phi_{m})\rbrace=1$ and $\mathbb{E}\lbrace\sin(2\phi_{m})\rbrace=0$. Thus, according to (\ref{mu_cos_D}), the real and imaginary parts of $\widetilde{H}_{DI}$ can be expressed as
%\vspace{-2mm}
\begin{equation}
\begin{aligned}
	&\Re{(\tilde{H}_{DI})} \sim N (0,\frac{M}{2}+\frac{1}{2}\sum_{m_y=1}^{M_y} \sum_{m_x=1}^{M_x} \cos(2\alpha)) \ ,\\
& \Im{(\tilde{H}_{DI})} \sim N (0,\frac{M}{2}-\frac{1}{2}\sum_{m_y=1}^{M_y} \sum_{m_x=1}^{M_x} \cos(2\alpha)) \ .
\end{aligned}
\label{discrete_channel}
\end{equation}
The real and imaginary parts of $\widetilde{H}_{DI}$ are correlated, with covariance $\mathbf{R}_{DI}$ expressed as
%\vspace{-3mm}
\begin{equation}
\begin{aligned}
\mathbf{R}_{DI}&=\mathbb{E}\lbrace \sum_{m_y=1}^{M_y} \sum_{m_x=1}^{M_x} \cos(\phi_{m}+\alpha)\sin(\phi_{m}+\alpha) \rbrace \\
&= \sum_{m_y=1}^{M_y} \sum_{m_x=1}^{M_x} \frac{1}{2}\sin(2\alpha) \ .
\end{aligned}
\label{correlation}
\end{equation}
The real and imaginary parts are independent when $\mathbf{R}_{DI}=0$. Assume $m_x=m_y$ and $d_x=d_y$, which is a typical configuration of URA RIS. Then, $\mathbf{R}_{DI}=0$ when the following condition is satisfied  
\begin{equation}
	\gamma = \frac{b\pi}{2kd_x} \ , b \in \mathbb{Z} \ ,
	\label{Rlow}
\end{equation}
\vspace{-2mm}
\begin{equation}
	\gamma = \cos\psi_{i} \sin\theta_{i} + \cos\psi_{o} \sin\theta_{o} +\sin\psi_{i} \sin\theta_{i}+\sin\psi_{o} \sin\theta_{o} \ .
\end{equation}

Additionally, the reflection channel variance is $\sigma^2_{DI} \leq M$. The less-equal sign is due to the covariance between the real and imaginary parts. When $\mathbf{R}_{DI} = 0$, 1-bit RIS provides the same variance. The magnitude and phase distributions for 1-bit DIPS can be turned into deriving the envelope and phase of correlated Gaussian quadratures according to \cite{4400760}. 

\vspace{-2mm}
\section{Secret Key Rate}
SKR is the upper bound of the information bits per channel sample generated between Alice and Bob. Since when Eve is several wavelengths away from Alice and Bob, its channel is considered uncorrelated with the channel between Alice and Bob, which is tested in practical experiments \cite{2017arXiv170308239S}. The SKR can be expressed as
\begin{equation}
	R_s=I(y_A;y_B| y_E) = I(y_A;y_B) \ ,
\end{equation}
where $y_A$, $y_B$, and $y_E$ are the channels estimated by Alice, Bob, and Eve, respectively. 

\vspace{-2mm}
\subsection{CIPS, CGPS, and $B \geq 2$ DIPS}
\label{con_SKR_section}
The mutual information between $y_A$ and $y_B$ is determined by their variances. The direct channel adds a mean to the reflection channel, without influencing the variance. Therefore, based on the channel expressions in (\ref{received_signal}) and Section \ref{continuous}, the SKR $R_s$ for CIPS and $B\geq2$ DIPS can be derived as
\begin{equation}
	R_{s,CI} = \log_2 (1 + \frac{M/2}{2\sigma_z^2+\frac{2\sigma_z^4}{M}}) \ ,
	\label{SKR_con}
	\vspace{-3mm}
\end{equation}
where $\sigma_z^2$ is the noise power.

The SKR for CGPS can be calculated as (\ref{SKR_con}) by substituting $\sigma^2_{CI}=M$ by $\sigma^2_{CG}=Nq^2$.

\begin{figure*}[htb]
	\centering
	\hspace{-9mm}
	\subfigure[Variance when RIS has different element number.]
	{
		\includegraphics[scale=0.44]{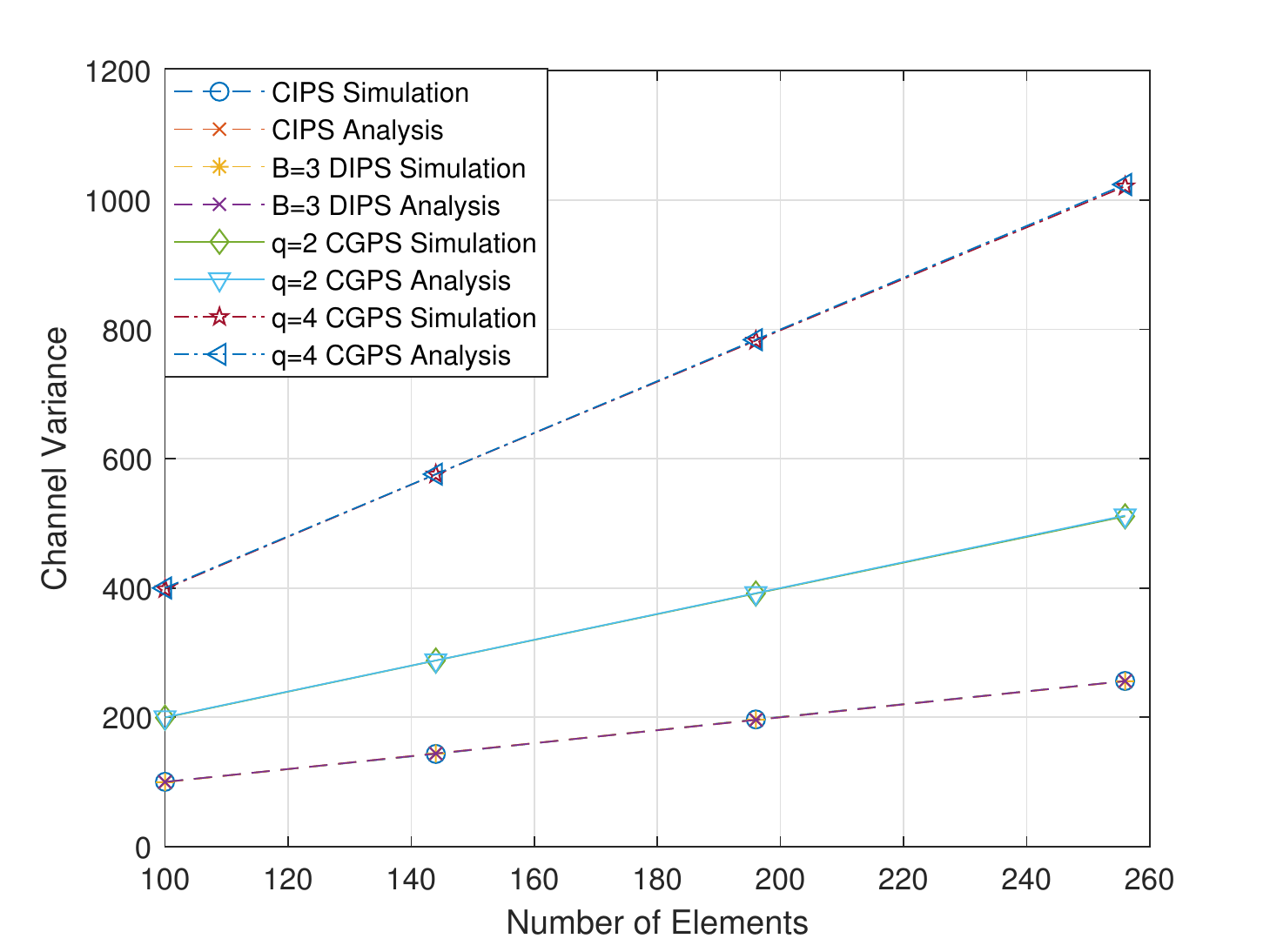}
		\label{figure:var_con_dis3}
	}
	\hspace{-9mm}
	\subfigure[Magnitude distribution.] 
	{
		\includegraphics[scale=0.44]{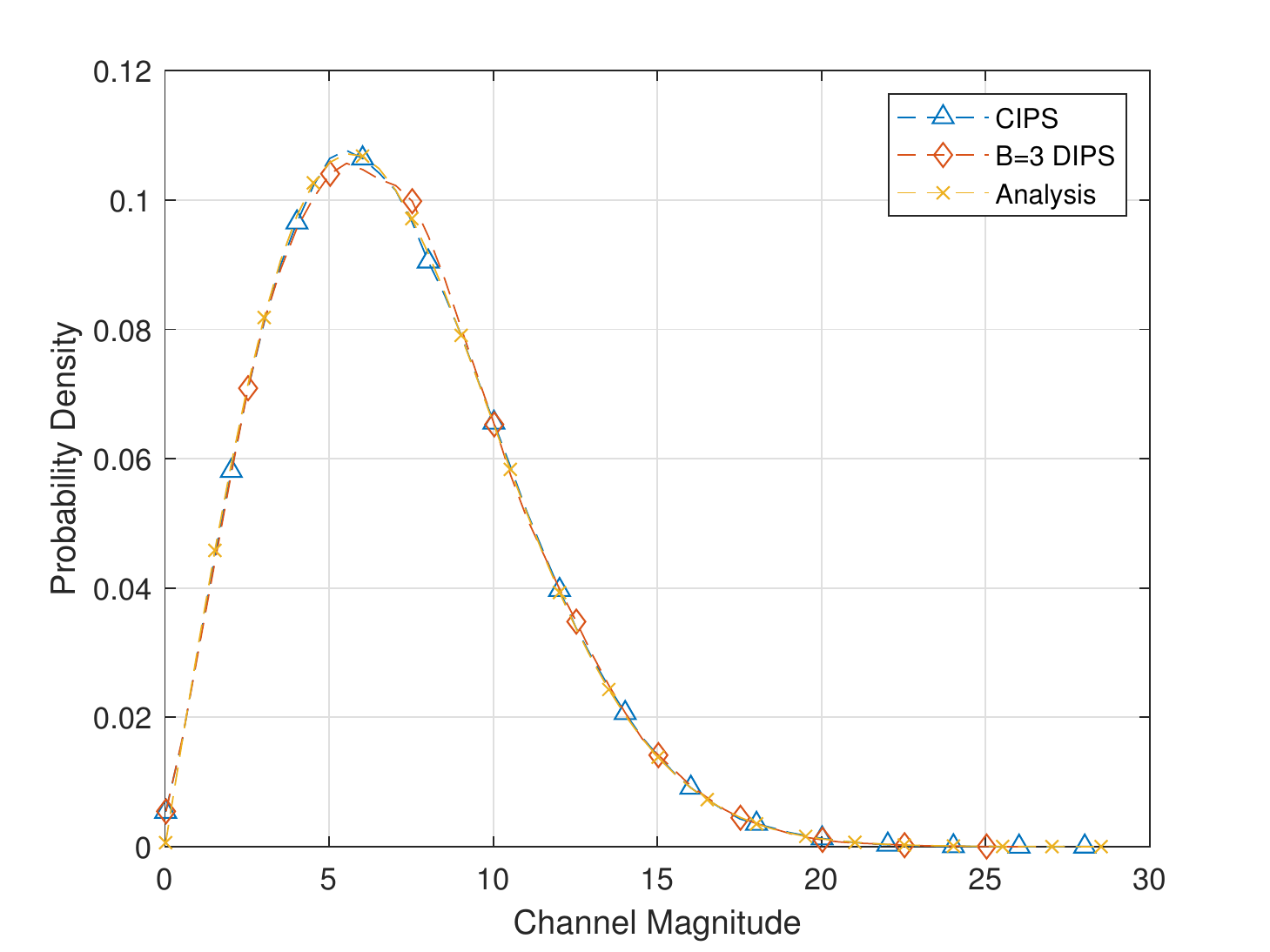}
		\label{figure:mag_con_dis3}
	}
	\hspace{-8mm}
	\subfigure[Phase distribution.] 
	{
		\includegraphics[scale=0.44]{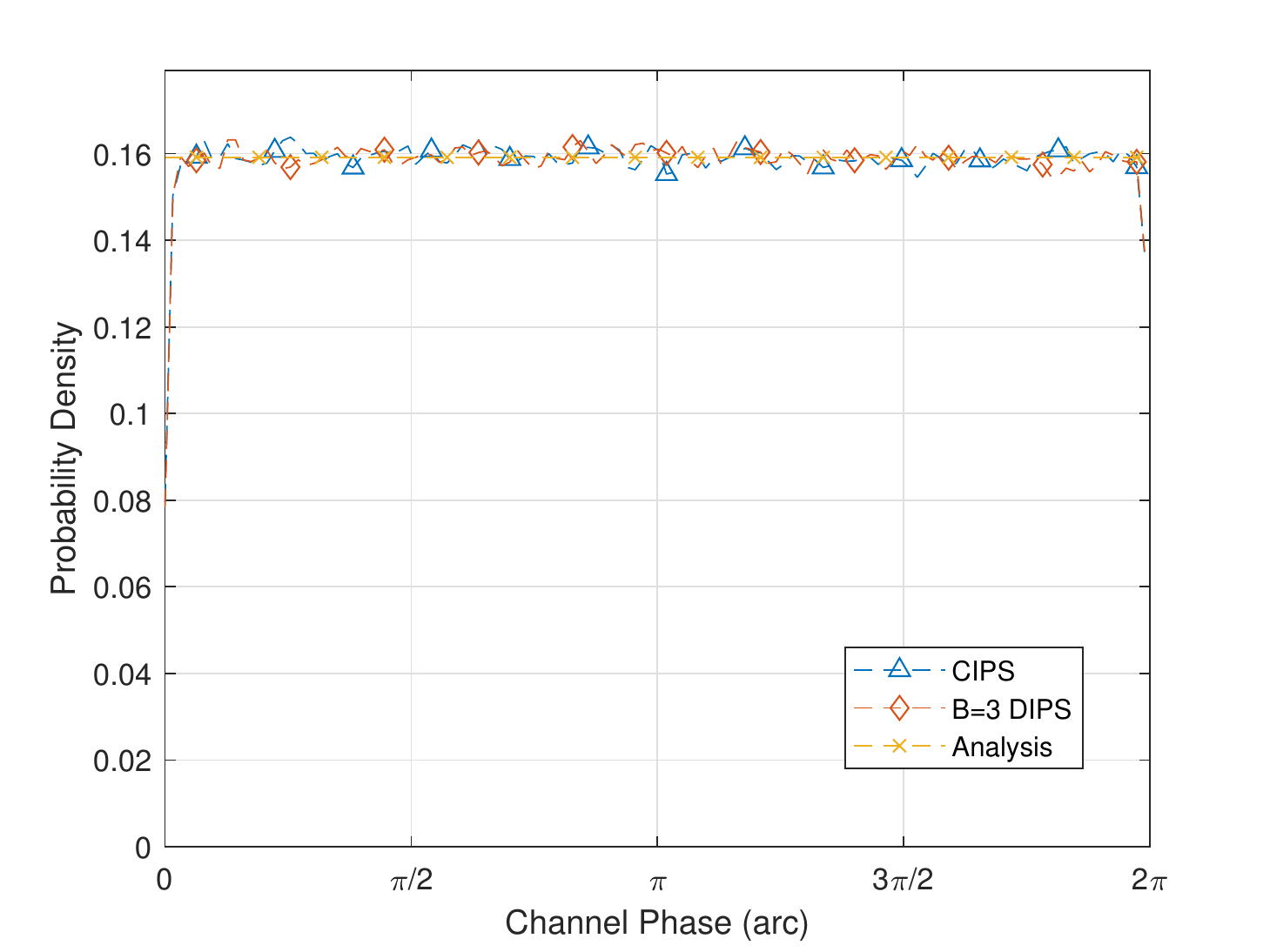}
		\label{figure:phase_con_dis3}
	}
	\hspace{-9mm}
	\caption{The channel variance of CIPS, $q=2,4$ CGPS, and $B=3$ DIPS. The magnitude distribution and phase distribution of the CIPS and $B=3$ DIPS.}
	\vspace{-2mm}
\end{figure*}

\subsection{$B=1$ DIPS}
The estimated SKR under $B=1$ DIPS can be expressed as
\begin{equation}
\begin{aligned}
R_{s,DI} &  \leq \lbrace I(\Re{(y_A)};\Re{(y_B)}) + I(\Im{(y_A)};\Im{(y_B)}) \rbrace \\ 
&= \log_2 \lbrace (1\!+\!\frac{\sigma^2_{real,DI}}{2\sigma_z^2+\frac{\sigma_z^4}{\sigma^2_{real,DI}}}) (1\!+\!\frac{\sigma^2_{imag,DI}}{2\sigma_z^2+\frac{\sigma_z^4}{\sigma^2_{imag,DI}}}) \rbrace,
\end{aligned}	
\label{SKR_discrete}
\end{equation}
%\overset{\text{(a)}}{\leq} 
where $\sigma^2_{real,D,I}$ and $\sigma^2_{imag,D,I}$ are the variances of the real and imaginary parts in (\ref{discrete_channel}). The less-than sign is because in (\ref{correlation}), the 1-bit DIPS RIS channel real and imaginary parts are correlated. The sign takes eqivalence when the real and imaginary parts are independent taking the condition in (\ref{Rlow}). 

\begin{figure*}[htb]
	\centering
	\hspace{-9mm}
	\subfigure[Variance when RIS has different element number.]
	{
		\includegraphics[scale=0.44]{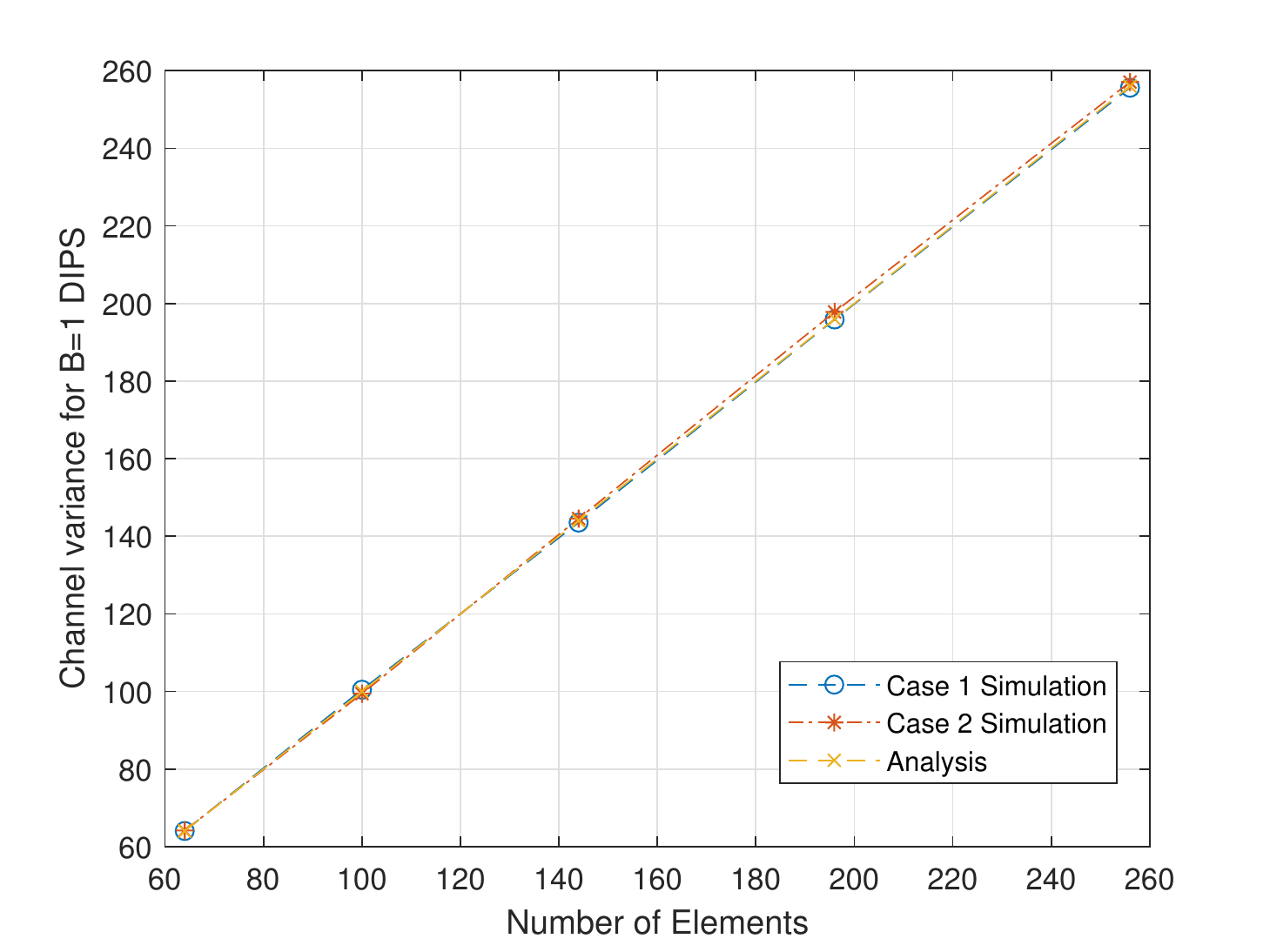}
		\label{figure:var_1bit}
	}
	\hspace{-9mm}
	\subfigure[Real part distribution.] 
	{
		\includegraphics[scale=0.44]{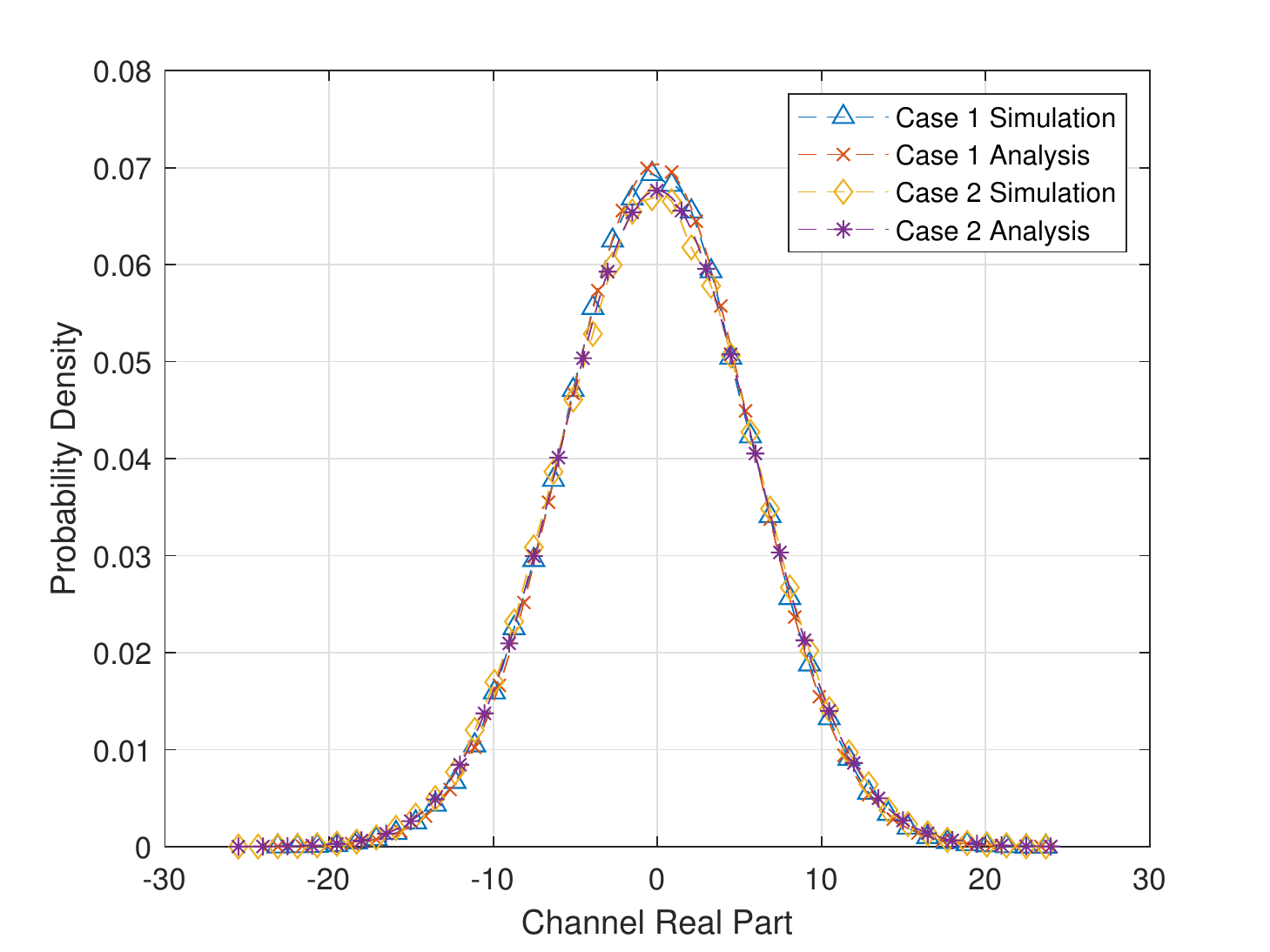}
		\label{figure:1bit_real}
	}
	\hspace{-8mm}
	\subfigure[Imaginary part distribution.] 
	{
		\includegraphics[scale=0.44]{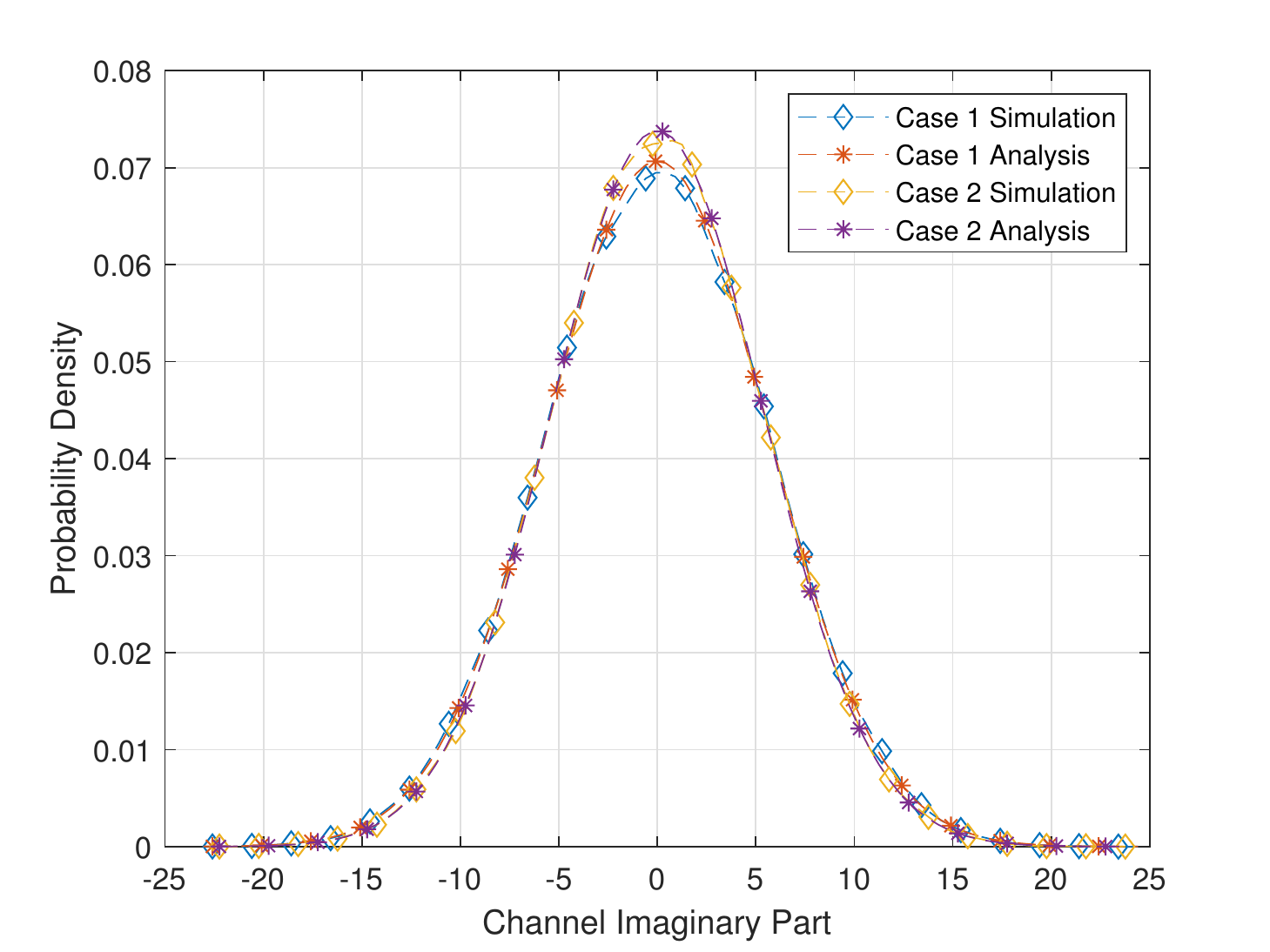}
		\label{figure:1bit_imag}
	}
	\hspace{-9mm}
	\caption{The variance, real part distribution and imaginary part distribution of the channel assisted by $B=1$ DIPS RIS.}
		\vspace{-4mm}
\end{figure*}
\section{Simulation Results}\label{Simulation Results}
In this section, the analytical channel distributions, channel parameters, and SKR for CIPS, CGPS, and DIPS are verified by simulations.
\subsection{Verification for CIPS and CGPS Channel}
\label{continuous_simulation}
Element spacing is set to $d=\frac{\lambda}{2}$. Incident and reflection angles are set as $\bm{\Omega_{i}}=(30^{\circ}, 30^{\circ})$, and $\bm{\Omega_{o}}=(150^{\circ},60^{\circ})$. For CGPS, set group size $q=2,4$.
\subsubsection{Channel Variance}
The simulation for the reflection channel variance with respect to $M$ is plotted in Fig. \ref{figure:var_con_dis3}. The simulation results match analytical results in (\ref{var_patt}) (\ref{var_patt_group}). The variance means the RIS has a great potential to induce channel variation since it is usually implemented with a huge number of elements. The variance for the CGPS is larger than the CIPS, and $q=4$ CGPS variance is larger than $q=2$ CGPS because the variance increases quadratically with $q$.

\subsubsection{Magnitude and Phase Distributions}
Set $M=64$, with $M_x=M_y=8$. The magnitude and phase distributions for CIPS are presented in Fig. \ref{figure:mag_con_dis3} and Fig. \ref{figure:phase_con_dis3}, respectively. The magnitude and phase distribution simulation results match the analytical results. The magnitude and phase are a Rayleigh distribution and a uniform distribution being independent of $\bm{\Omega}_{i}$ and $\bm{\Omega}_{o}$, respectively. The independence of $\bm{\Omega}_{i}$ and $\bm{\Omega}_{o}$ means that whatever angles Alice and Bob locate in, the channel randomness remains the same. Note that the sharp drops around 0 and $2\pi$ in Fig. \ref{figure:phase_con_dis3} result from several adjacent probability density values outside the bound $[0,2 \pi)$, being equal to 0, are averaged to plot a smoother p.d.f. Additionally, the channel distribution Eve receives stays unchanged, and no more information leakage dependent on angles. Note that if Eve locates at the same angle as Bob\footnote{The situation Eve locates at the same angle as Bob do not happen in most cases.}, the information is fully leaked. This is because, in the far-field environment, Eve and Bob are differentiated by their locating angles. The uniform phase distribution means that the values in $[0,2\pi)$ are taken with equal probability. Thus, when CSI is utilized to generate keys, it greatly prevents Eve from guessing the particular phase of the channel between Alice and Bob.

\subsection{Verification for DIPS}
\subsubsection{RIS Weight $B\geq 2$}
Set $B=3$. Other parameters remain the same as the continuous case. The simulations for the variance, magnitude, and phase of the reflection channel are plotted in Fig. \ref{figure:var_con_dis3}, Fig. \ref{figure:mag_con_dis3}, and Fig. \ref{figure:phase_con_dis3}. The simulation results are the same as the ideal CIPS. This means that the ideal channel distribution and uniform phase distribution can be achieved in a more practical situation. 

\subsubsection{RIS Weight $B = 1$}
Set $B=1$, and two pairs of input and output angles, with case 1:  $\bm{\Omega_{i}}=(30^{\circ},30^{\circ})$, $\bm{\Omega_{o}}=(150^{\circ}, 60^{\circ})$, and case 2: $\bm{\Omega_{i}}=(110^{\circ}, 50^{\circ})$, $\bm{\Omega_{o}}= (310^{\circ}, 20^{\circ})$. Other parameters are the same as continuous case. The simulation for the reflection channel variance is plotted in Fig. \ref{figure:var_1bit}, where the variances increase linearly with respect to $M$. 

The real and imaginary part of the reflection channel for two cases when $M_x=M_y=8$ are plotted in Figure. \ref{figure:1bit_real} and Figure. \ref{figure:1bit_imag}, respectively. The simulation results match the analytical result in (\ref{discrete_channel}) well. the real and imaginary part distributions are dependent on values of $\bm{\Omega}_{i}$ and $\bm{\Omega}_{o}$. 

\subsection{Secret Key Rate}
\begin{figure}[htb]
	\centering
	\includegraphics[scale=0.6]{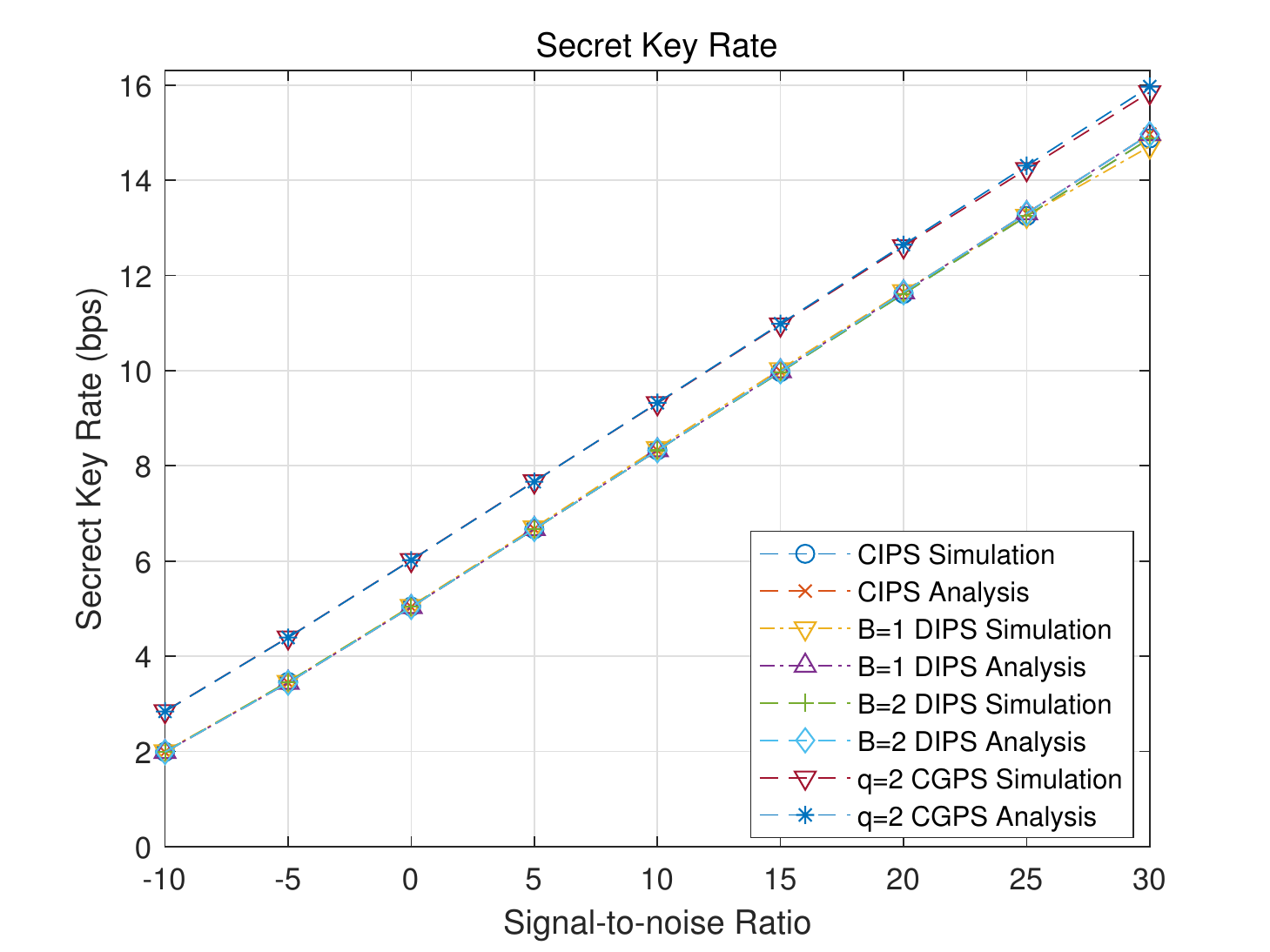}
	\caption{Secret key rate when RIS is CIPS, $q=2$ CGPS, $B=2$ DIPS, and $B=1$ DIPS.}
	\label{MI}
\end{figure}

The RIS element number is set to $M=64$. The \emph{Information Theoretical Estimators (ITE) Toolbox} is used to simulate the mutual information \cite{szabo14information}. The SKR for CIPS, $q=2$ CGPS, $B=2$ DIPS, and $B=1$ DIPS against signal-to-noise ratio (SNR) are plotted in Fig. \ref{MI}. The simulation results match the analytical results in (\ref{SKR_con}) and (\ref{SKR_discrete}). The CGPS SKR is larger than the individual shifting. 
\vspace{-2mm}

\vspace{-4mm}
\section{Conclusions}\label{Conclusions}
In this paper, we utilize RIS for mmWave physical layer security secret key generation, and completes the underlying mathematical principles by deriving the channel distribution. Though mmWave static channels lack randomness and multipath, the RIS can induce an artificial Rayleigh/Rician fading randomness, without adding transceiver costs. Specifically, based on the RIS-assisted secret key generation model, we consider RIS weights have CIPS, CGPS, and DIPS to produce artificial channel randomness. We derive the channel distribution and its parameters. The CGPS is able to produce more randomness at legal parties Alice and Bob, compared to individual shifting. The DIPS channel distribution is dependent on the quantization bit. The SKR for the above RIS settings is derived to evaluate the performance. Our theoretical conclusions are verified by simulations. While this work focuses more on the channel between Alice and Bob, the detailed analysis of leakage to eavesdroppers and DGPS will be our future work.

\vspace{-2mm}

\bibliography{conference}
\bibliographystyle{IEEEtran}

\end{document}